\documentclass[preprint,prl,showpacs,byrevtex,superscriptaddress]{revtex4}
\usepackage{graphicx}
\usepackage{epsfig}

\begin{document}

\title{Theory of Microwave Parametric Down Conversion and Squeezing Using Circuit QED}
\author{K.~Moon}
\affiliation{Sloane Physics Laboratory, PO Box 208120, Yale
University, New Haven, CT 06520-8120}
\affiliation{Department of Physics and Institute of Physics and
Applied Physics, Yonsei University, Seoul 120-749, Korea}
\author{S.~M.~Girvin}
\affiliation{Sloane Physics Laboratory, PO Box 208120, Yale
University, New Haven, CT 06520-8120}

\date{\today}

\begin{abstract}
We study theoretically the parametric down conversion and squeezing
of microwaves using cavity quantum electrodynamics of a
superconducting Cooper pair box (CPB) qubit located inside a
transmission line resonator. The non-linear susceptibility $\chi_2$
describing three-wave mixing can be tuned by dc gate voltage applied
to the CPB and vanishes by symmetry at the charge degeneracy point.
We show that the coherent coupling of different cavity modes through
the qubit can generate a squeezed state. Based on parameters
realized in recent successful circuit QED experiments, squeezing of
$95\%\sim 13{\rm dB}$ below the vacuum noise level should be readily
achievable.

\end{abstract}
\maketitle
%Introduction
Squeezed states are a valuable tool to usefully manipulate the
Heisenberg uncertainty principle by reducing the quantum
fluctuations of a certain variable of interest at the expense of
increased uncertainty for its conjugate variable. Using squeezed
states one can perform very quiet measurements much below the vacuum
noise level \cite{Kimble}. Squeezed states manifest the quantum
coherent nature of light and provide a chance to beat the standard
quantum limit by preferentially doing an experiment using the
squeezed quadrature alone \cite{WallMilburn,Drummond}. Squeezed
states have been experimentally observed in a nonlinear optical
cavity experiment \cite{First,OPO}. Recently, the theory of
squeezing in a high-$Q$ cavity was considered \cite{Almeida}. Upon
the injection of a high-energy photon, a nonlinear optical medium
can coherently generate two photons, the sum of whose frequencies is
equal to that of the high-energy photon via optical parametric down
conversion (PDC). If one injects low-energy photons instead, one may
induce second harmonic generation, which also forms a squeezed
state. In addition to this three-wave mixing, four-wave mixing can
be used to generate squeezed states. In pioneering condensed matter
experiments, the Josephson junction parametric amplifier was used in
the microwave regime to produce $(47\pm 8) \%\sim 3 {\rm dB}$
squeezing below the vacuum level \cite{Yurke1,Yurke2,Yurke3} via
degenerate four wave mixing. There has been tremendous recent
progress in realizing quantum optics physics in electrical circuits.
It is now possible to experimentally reach the extreme strong
coupling limit of cavity QED \cite{Blais,Nature,Schuster} and to see
very strong microwave non-linearities in high inductance small scale
Josephson junctions whose Hamiltonian can be controlled with
remarkable accuracy \cite{JBA}. Coherent dynamics of a flux qubit
coupled to a harmonic oscillator in a SQUID circuit has also been
demonstrated \cite{Delft}.

\begin{figure}
\epsfxsize=5.in \epsfysize=2.in \epsffile{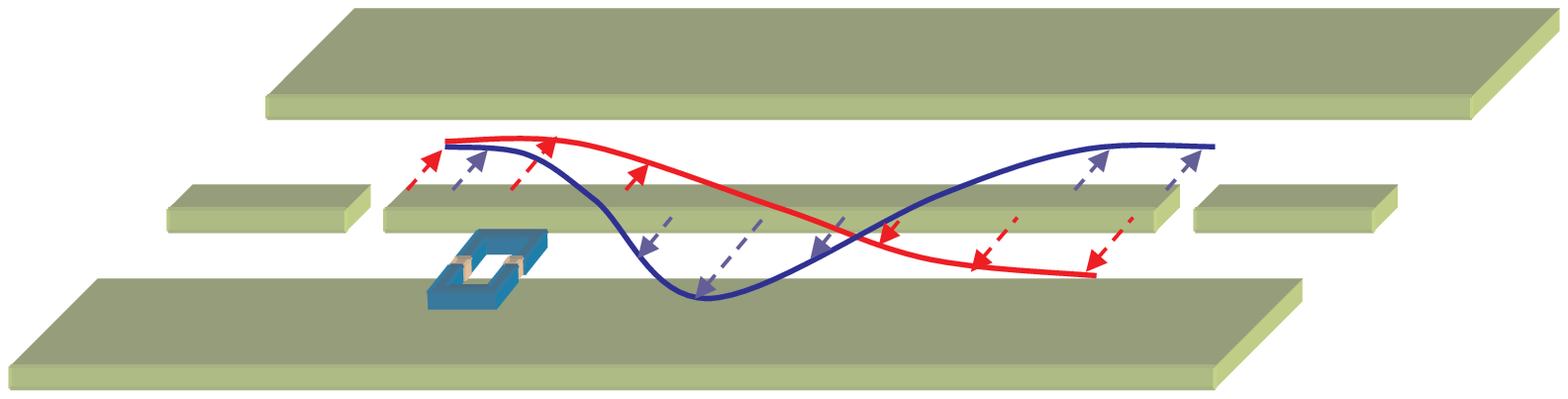}
\caption{(Color online) The schematic diagram for the circuit QED
coupled to a Cooper-pair box located at one edge of the cavity.
The red and blue lines represent the fundamental and the second
harmonic cavity modes, respectively. The change of CPB position
from the edge will vary the coupling strengths $g_1$ and $g_2$.}
\label{cQED}
\end{figure}

Motivated by these experimental advances in `circuit QED', we here
study squeezing in a system consisting of a Cooper pair box (CPB)
located inside a high $Q$ coplanar waveguide resonator. Two
different discrete photon modes (fundamental and first harmonic) are
coupled through the CPB  as shown in Fig.~\ref{cQED} and squeezing
occurs via three-wave mixing. Compared to the cavity QED of atomic
physics, circuit QED has the advantages of infinite transit time of
the `atom' (qubit) inside the cavity and much stronger coupling
between qubit and photon. We emphasize that strong coupling means we
only need a single `atom'. The circuit QED system recently
successfully demonstrated strong coupling (vacuum Rabi splitting)
between a single photon and a qubit in all solid state system
\cite{Nature}. While this architecture offers a very quiet
environment leading to excellent coherence of the qubit ($T_1\sim
7\mu{\rm s}$, $T_2^*\sim 500{\rm ns}$) \cite{Schuster}, the
solid-state environment still leads to qubit decay rates and
dephasing rates much larger than those in corresponding atomic
physics microwave cavity QED experiments \cite{haroche}. Hence it is
crucial to investigate these environmental effects on the efficiency
of squeezing. Using both numerical and analytical calculations based
on the currently available experimental parameters, we have
estimated that the circuit QED system can readily produce about $95
\%$ squeezing, that is, $13 {\rm dB}$ below the vacuum noise level.

%Squeezing for 1/T_1=0 and 1/T_2=0.
We start with the following Hamiltonian to describe the coupled
system of qubit and cavity photons for microwave circuit
QED\cite{Blais,Nature}: $H = H_0 + H_I$. The Hamiltonian $H_0$ is
given by
\begin{equation}
H_0={E_{01} \over 2}\sigma_z + \hbar\omega_1 a_1^{\dagger} a_1 +
\hbar\omega_2 a_2^{\dagger} a_2 , \label{Hamiltonian}
\end{equation}
where $E_{01}=\sqrt{E_J^2+E_{\rm el}^2}$, $E_J$ is the Josephson
coupling energy, $E_{\rm el}=4E_C(1-2N_g)$, $E_c$ the charging
energy, $N_g$ the gate charge, and $\omega_1$ the angular frequency
of the fundamental resonator mode and $\omega_2 =2\omega_1$ is the
first harmonic frequency. The coupling Hamiltonian $H_I$ can be
written
\begin{equation}
H_I =-\left[g_1(a_1^{\dagger}+a_1)+g_2
(a_2^{\dagger}+a_2)\right]\left[1-2N_g + \sin\theta \,\sigma_x -
\cos\theta \,\sigma_z\right], \label{Coupling}
\end{equation}
where $g_i$ represents the coupling strength between the qubit and
the $i$th cavity photon mode, and $\theta=\tan^{-1}(E_J/E_{\rm
el})$. At the charge degeneracy point (CDP), that is, $N_g = 1/2$,
$H_I$ reduces to the Jaynes-Cummings Hamitonian, since
$\cos\theta\cong -(4E_C/E_J)(1-2N_g)=0$ and $\sin\theta= 1$. One can
see that away from the CDP, we have couplings other than the
Jaynes-Cummings term, whose strength linearly increases with the
deviation. Because the only relevant cavity frequencies will be the
fundamental and first harmonic, we expect degenerate PDC to occur in
our system, where a single high-energy photon coherently generates
two photons, each with half the frequency. We emphasize that this is
degenerate PDC because the cavity only has discrete modes. In order
to achieve squeezing via PDC in circuit QED, the system should be
able to convert a single $\omega_2$ photon into two $\omega_1$
photons, which requires a term of the form $a_1^\dagger a_1^\dagger
a_2$ in the effective Hamiltonian. This will result from the third
order processes in terms of $H_I$, which will be in general
negligible since $g_1^2 g_2/\omega_1^3 \ll 1$ in typical
experiments\cite{Nature}.  However we can resonantly enhance the
process by tuning either $\omega_1$ or $\omega_2$ close to $E_{01}$.
We have chosen the case $\omega_2 \cong E_{01}$.

We first apply the following unitary transformation $U_1$ to the
Hamiltonian $H$
\begin{equation}
U_1=\exp\left[\sum_{i=1,2} {g_i \over
\omega_i}\left(a_i-a_i^{\dagger}\right)\left(1-2N_g - \cos\theta
\,\sigma_z\right)\right].
\end{equation}
This corresponds to shifting the centers of the harmonic oscillator
coordinates $X_i=a_i+a_i^{\dagger}$ by $(2g_i/ \omega_i)(1-2N_g -
\cos\theta \,\sigma_z)$. Subsequently, we apply the unitary
transformation
\begin{equation}
U_2=\exp\left[{g_1 \sin\theta \over
2(E_{01}-\omega_1)}\left(a_1^{\dagger}\sigma^--a_1\sigma^+\right)\right].
\end{equation}
Upon application of these two unitary transformations, we obtain
after perturbative expansion in $g_1$ and $g_2$, the following
Hamiltonian: $H_{\rm eff}=U_2 U_1 H U_1^\dagger U_2^\dagger=H_0 +
H^\prime$ for $E_{01}\cong \omega_2$,
\begin{equation}
H^\prime=- {1\over 2} g_2 \sin\theta (a_2^{\dagger}\sigma^- +
a_2\sigma^+)+{g_1^2 \sin 2\theta \over
2\omega_1}(a_1^{\dagger}a_1^{\dagger}\sigma^- + a_1 a_1 \sigma^+
). \label{EffHamiltonian1}
\end{equation}
We define the energy detuning between the cavity photon frequency
$\omega_2$ and the qubit energy splitting $E_{01}$ to be
$\Delta\equiv E_{01} -\omega_2$ and consider the case $g_2 \ll
\Delta\ll \omega_1$. It is this Hamiltonian which we will study
numerically.  However to develop an analytical understanding, we can
apply the following additional unitary transformation $U_3$
\begin{equation}
U_3=\exp\left[{g_2 \sin\theta \over
2\Delta}\left(a_2^{\dagger}\sigma^--a_2\sigma^+\right)\right].
\end{equation}
Finally, we obtain the following low-energy effective Hamiltonian
through ${\hat H}_{\rm eff}=U_3 H_{\rm eff} U_3^\dagger$,
\begin{equation}
{\hat H}_{\rm eff}= H_0 + {g_2^2 \sin^2\theta\over 2\Delta}
\left[1 + \sigma_z (2n_2 +1)\right] + {\zeta\over 2}
(a_1^{\dagger}a_1^{\dagger}a_2 + a_1 a_1
a_2^{\dagger})\sigma_z,\label{EffHamiltonian2}
\end{equation}
where the second term on the right represents the Lamb and light
shifts of the qubit splitting frequency and $\zeta = (2 g_1^2 g_2
\sin\theta \sin 2\theta/\Delta\omega_1)$. The third term is the
desired squeezing term. ${\hat H}_{\rm eff}$ is exactly the standard
Hamiltonian for degenerate optical PDC including the Lamb
shift\cite{WallMilburn}. Note that the squeezing operator couples to
the qubit state $\sigma_z$ and the phase of the squeezed quadrature
will shift by $\pi/2$ if the qubit is placed in the excited state.

In deriving the above result, we have neglected the effect of cavity
damping $\sum_{i=1,2}\sqrt{\kappa_i}\int_{-\infty}^\infty d\omega\,
a_i b_\omega^\dagger$, where $b_\omega$ represents the continuum
modes outside the cavity and $\kappa_i$ the cavity loss rates.
Unitary transformation of the damping terms leads to radiative atom
damping \cite{Blais} and two-photon decay terms \cite{Yurke2P} such
as $(\sqrt{\kappa_2}\zeta/2\Delta)\int_{-\infty}^\infty d\omega\,
a_1 a_1 b_\omega^\dagger$. The two-photon decay rate is much smaller
than the cavity loss $\kappa_2$ by a factor of $(\zeta/\Delta)^2/2$,
and hence we will neglect it.  Similarly the radiative atom damping
term is smaller than the intrinsic atom decay rate and we will
neglect it. In the limit of weak pumping, down converted pairs of
photons are produced incoherently at a rate  ${\bar n}_2
(\zeta/\kappa_1)^2$ given by Fermi's golden rule  with ${\bar n}_2$
the average number of pump photons.

To understand the squeezing produced by strong pumping, one needs to
consider the substantial de-excitation rate $\gamma$ from the
excited state of solid-state qubits. Furthermore, the qubit
dephasing rate $\gamma_\varphi$ is typically at least one order of
magnitude greater than $\gamma$. For charge qubits, optimal phase
coherence occurs at the charge degeneracy point \cite{Saclay} where
(it happens that) the symmetry prevents the three wave mixing which
we require for PDC. Hence it is crucial to take into account these
effects to obtain a realistic estimate of squeezing. For a deviation
of 10\% from the charge degeneracy point, coherence times of
$T_2\sim 200$ns have been demonstrated \cite{Saclay}. We start with
the Hamiltonian ${\tilde H}_I$ obtained via the unitary
transformation $U_1 U_2$ in Eq.~\ref{EffHamiltonian1}, which is
defined in the rotating wave frame of $\omega_2$,
\begin{equation}
{\tilde H}_I={\Delta \over 2}\sigma_z + g_2 \sin\theta
|\alpha_p|\sigma_y +{\Gamma \over
4}(a_1^{\dagger}a_1^{\dagger}\sigma^- + a_1 a_1 \sigma^+ ).
\label{RWHamiltonian}
\end{equation}
Here $\Gamma=2g_1^2 \sin 2\theta/\omega_1$ and we have taken the
pump to be classical: $\langle a_2\rangle=\alpha_p=i|\alpha_p|$.

\begin{figure}
\epsfxsize=5.in \epsfysize=4.in \epsffile{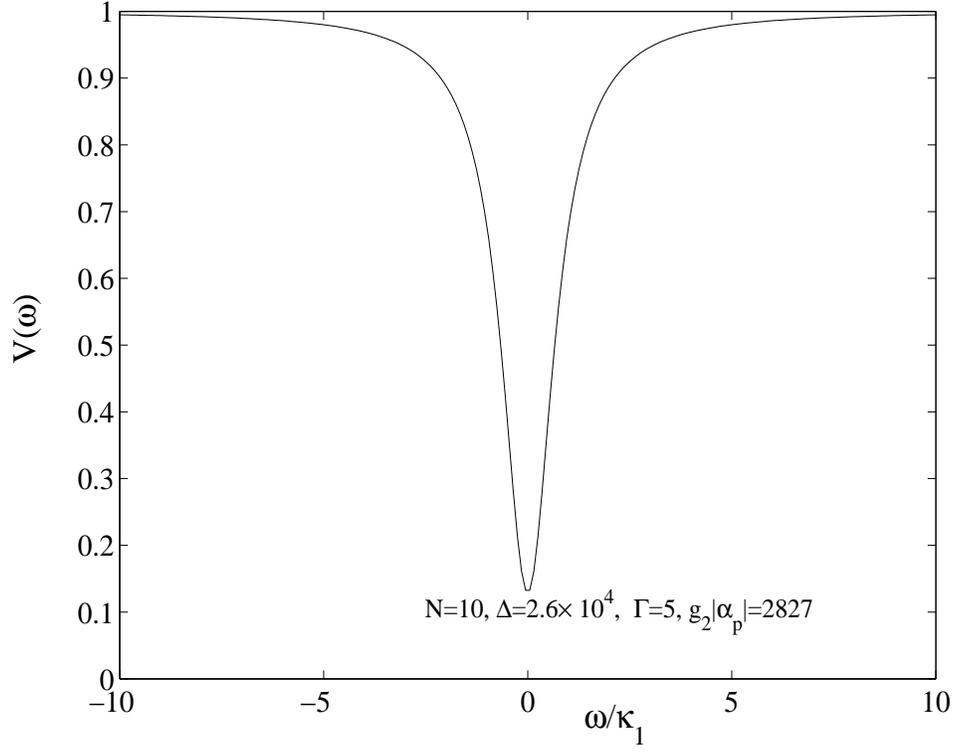}
\caption{ The quadrature variance $V(\omega)$ for $X_1$ is plotted
as a function of $\omega$ for $N=10, \Delta=2.6\times 10^4,
\Gamma=5, g_2|\alpha_p|=2827$ in units of $\kappa_1$. The maximum
squeezing is obtained at $\omega=0$.} \label{Qsim1}
\end{figure}

\begin{figure}
\epsfxsize=5.in \epsfysize=4.in \epsffile{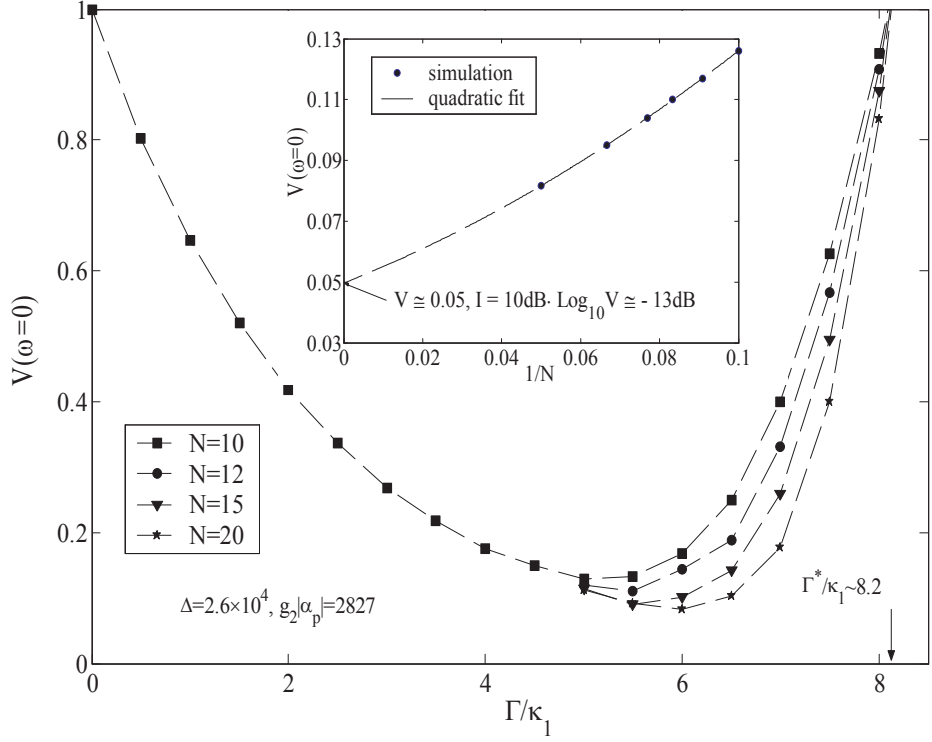}
\caption{ The quadrature variance $V(\omega=0)$ for $X_1$ is
plotted as a function of $\Gamma$ for several different values of
$N=10,12,15,20$. The minimum value of $V(\omega=0)$ decreases with
the increase of the maximum photon number $N$. The critical value
of $\Gamma$ is about $8.2$ in units of $\kappa_1$. In the inset,
the maximum output squeezing for several values of $N$ is plotted
with respect to $1/N$. By extrapolation to $N=\infty$, we obtain
the $V_{\rm min}(\omega=0)=0.05$, that is, $-13{\rm dB}$.}
\label{Qsim2}
\end{figure}

Following the standard quantum theory of damping, we investigate the
coupled system of qubit and cavity plus the reservoir. After
integrating out the reservoir degrees of freedom and using the
Markov approximation, one obtains the master
equation\cite{WallMilburn} for the reduced density matrix $\rho$ of
the qubit plus cavity system,
\begin{equation}
{d\rho \over dt}= -i[{\tilde H}_I, \rho] + \kappa_1 \left[a_1 \rho
a_1^\dagger - {1\over 2} a_1^\dagger a_1 \rho - {1\over 2}\rho
a_1^\dagger a_1 \right] + {\gamma\over 4} \left[\sigma^- \rho
\sigma^\dagger - {1\over 2} \sigma^\dagger \sigma^- \rho - {1\over
2}\rho \sigma^\dagger \sigma^-\right] + {\gamma_\varphi \over
2}\left[\sigma_z \rho \sigma_z -\rho\right]. \label{Density}
\end{equation}
Using the quantum regression theorem\cite{WallMilburn},
 the variances $V(\omega)$
(homodyne spectrum) of quadrature $X_1= a + a^\dagger$ and $X_2= (a
- a^\dagger)/i$ for the {\em output} cavity photon mode are given by
\begin{equation}
V(\omega)= 1 \pm \kappa_1 \int_{-\infty}^\infty d\tau
e^{-i\omega\tau} {\rm Tr} \left\{ (a_1 \pm a_1^\dagger) e^{{\cal
L}\tau} \left(a_1\rho_{ss}\pm \rho_{ss}
a_1^\dagger\right)\right\}, \label{Noise-spectrum}
\end{equation}
where ${\cal L}$ is the Liouvillean operator, $\rho_{ss}$ the
density matrix at the steady state, and the ($+$) and ($-$) sign
correspond to $X_1$ and $X_2$ quadratures, respectively. We have
numerically calculated the $V(\omega)$ of quadrature $X_1$ and
$X_2$ based on an exact diagonalization study in the Hilbert space
of size $2N\times 2N$ corresponding to the two possible spin
states and the number of photons being restricted to less than
$N$.
 We have chosen the following conservative set of
experimental parameters:
%
% SMG0601 put in the correct frequency
%
$\omega_1/2\pi=3{\rm GHz}$, $T_1 = 2 {\rm \mu s}, T_2 = 100 {\rm
ns}, Q=5\times 10^5,\, g_2/2\pi=18 {\rm MHz}$, $E_C/E_J\cong 1$,
and $\sin\theta\cong 1$ near the CDP, where
$T_2=\gamma_\perp^{-1}$, $\gamma_\perp=\gamma_\varphi+\gamma/2$,
and $\Delta/\omega_1 \cong 0.05$\cite{Nature}. In
Fig.~\ref{Qsim1}, we have plotted $V(\omega)$ as a function of
$\omega$ for $\Delta=2.6\times 10^4, \Gamma=5, N=10$, and
$g_2|\alpha_p|=2827$ in units of $\kappa_1= \omega_1/Q$. We find
that the maximum squeezing is obtained at $\omega=0$ as expected.
The frequency width of the squeezing spectrum is controlled by the
cavity width $\kappa_1$. The variable $\Gamma$ is tunable by
varying the qubit gate voltage. It is experimentally observed that
the dephasing time $T_2$ decreases very rapidly, as the qubit is
detuned from the CDP \cite{Saclay}. Hence we will restrict the
relative deviations from the CDP to be small $|2N_g-1| \le 0.1$,
where $T_2\ge 100$ns. This restriction yields a maximum value of
$\Gamma\cong (8g_2^2/\omega_1)(E_C/E_J)(1-2N_g) \sim 0.53{\rm
MHz}$.
In Fig.~\ref{Qsim2}, $V(\omega=0)$ is plotted as a function $\Gamma$
for $N=10,12,15,20$. We have checked that $\Delta X_1\cdot\Delta
X_2$ satisfies the minimum uncertainty bounds and closely approaches
the minimum uncertainty condition. The maximum output squeezing is
obtained by extrapolation of the numerical results for finite $N$ to
$N=\infty$ yielding $V(\omega=0)=0.05$, that is, $-13{\rm dB}$, as
shown in the inset of Fig.~\ref{Qsim2}.

We have performed extensive simulations by varying the sets of
parameters. Increasing $\Delta$
 consistently improves the
squeezing because it reduces the effect of dephasing of the qubit.
 We can analytically study the effect of finite qubit decay
time and dephasing on the squeezing as follows. Based on
perturbative analysis, we have demonstrated that the PDC rate is
given by $(g_2\Gamma \sin\theta/\Delta)\sigma_z$ as shown in
Eq.~\ref{EffHamiltonian2}. When concerned with the spin dynamics
of the qubit alone, we may neglect the coupling term between
cavity photon and the qubit, which is much smaller than the other
terms in Eq.~\ref{RWHamiltonian}. By neglecting the coupling, we
obtain the following steady state solution for the spin
polarization $\langle\sigma_z\rangle$: $ \langle \sigma_z \rangle
= -\gamma (\gamma_\perp^2 + \Delta^2)/(4g_2^2 |\alpha_p|^2
\sin^2\theta \gamma_\perp +\gamma (\gamma_\perp^2 + \Delta^2))$.
In the absence of pump photons, ($|\alpha_p|=0$), the qubit
correctly decays down to the ground state, that is $\langle
\sigma_z \rangle=-1$. For large detuning, the finite pump only
slightly quenches the spin polarization.

\begin{figure}
\epsfxsize=5.in \epsfysize=4.in \epsffile{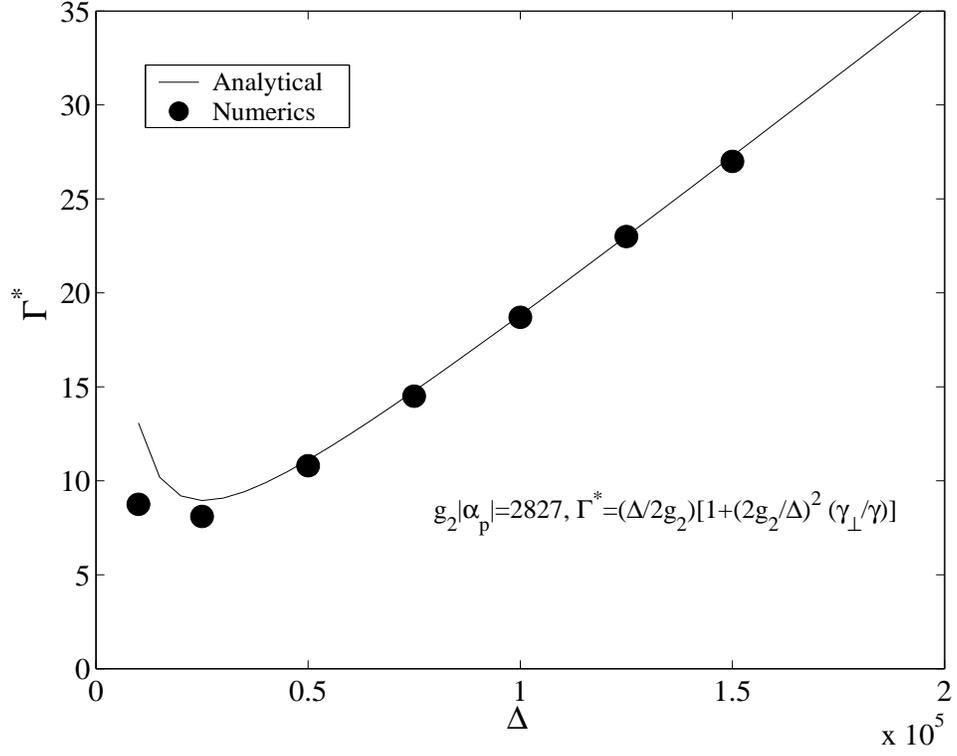}
\caption{ The critical value $\Gamma^*$ is plotted as a function
of $\Delta$. The dotted line represents the result from the
analytical formula and the filled circles from the numerical
simulation. Here $\Gamma^*$ and $\Delta$ are in units of
$\kappa_1$.} \label{Qsim4}
\end{figure}

Due to the quenching of qubit spin, the effective PDC parameter
$\chi_2$ is given by the following formula
\begin{equation}
\chi_2 = {g_2 \sin\theta |\alpha_p|\Gamma\over \Delta}{
\gamma_\perp^2 + \Delta^2 \over 4g_2^2 |\alpha_p|^2 \sin^2\theta
\left(\gamma_\perp/\gamma\right) + \gamma_\perp^2 + \Delta^2}.
\end{equation}
Maximum squeezing is achieved near the critical point $\chi_2 \cong
\kappa_1/2$\cite{WallMilburn}, which leads to the optimal value of
$\Gamma$ for $\Delta\gg \gamma_\perp$
\begin{equation}
\Gamma^* = {\kappa_1\over 2x}\left[ 1 + 4
\left(\gamma_\perp/\gamma\right) x^2\right],
\label{Ana}
\end{equation}
where $x=g_2 \sin\theta |\alpha_p|/\Delta$. In Fig.~\ref{Qsim4},
$\Gamma^*$ is plotted as a function of $\Delta$. The dotted line
represents the result from the analytical formula of Eq.~\ref{Ana}
which shows excellent agreement with the numerical simulation
(filled circles). By differentiating with respect to $x$, one can
obtain the minimum value of $\Gamma^*$ required to achieve the
maximum squeezing: $\Gamma_{\rm min}^*=2\kappa_1
(\gamma_\perp/\gamma)^{1/2}$ for $x^*=(\gamma/4\gamma_\perp)^{1/2}$.
Since the value of the $\Gamma/\kappa_1$ can reach as large as
$13.3$ for the chosen set of parameters, it is well above
$\Gamma_{\rm min}^*/\kappa_1=2 (\gamma_\perp/\gamma)^{1/2}\cong
8.9$. Hence the maximum squeezing can be realized with the chosen
experimental parameters.  The validity of perturbation theory in
$g_2$  \cite{Blais}
 imposes the following
constraint on the product of the pump amplitude and the coupling,
$g_2|\alpha_p| < (\Delta/2 \sin\theta)$. We note that one can
control the values of $g_1$ and $g_2$ independently by shifting the
position of the qubit within the cavity. When the CPB is located at
$1/4$ of the resonator length from the left edge, $g_2$ vanishes as
shown in Fig.~\ref{cQED}. Hence by placing the CPB slightly left of
the above position, the pump amplitude  can always be made large
enough for the classical approximation to be valid.

In summary, we have studied degenerate parametric down conversion
and squeezing in circuit QED, where a superconducting Cooper pair
box (CPB) qubit is located inside a transmission line resonator. We
have shown that away from the charge degeneracy point, the coherent
coupling of different cavity modes through the qubit can generate a
squeezed state via three-wave mixing. We have investigated the
effect of the finite qubit lifetime and dephasing on squeezing,
which will be crucial especially for the qubit away from the charge
degeneracy point. By performing both the numerical and analytical
calculations, we have demonstrated that the squeezing of about
$13{\rm dB}$ below the vacuum noise level can be obtained for the
currently available experimental parameters.

We thank A. Blais and J. Gambetta for valuable discussions and
guidance in the quantum optics simulations. This work was
supported in part by the National Security Agency (NSA) and
Advanced Research and Development Activity (ARDA) under Army
Research Office (ARO) contract numbers DAAD19-02-1-0045 and
 ARDA-ARO W911NF-05-1-0365, and by NSF ITR-0325580, NSF DMR-0342157 and
the Keck Foundation. K.M. wishes to acknowledge the financial
support of the LG Yonam Foundation and the National Program for
Tera-Level Nanodevices of the Korea Ministry of Science and
Technology as one of the 21 Century Frontier Programs.

\newpage

\clearpage

\newpage

\clearpage

\newpage

%\clearpage

%\newpage
%\begin{figure}
%\epsfxsize=5.in \epsfysize=4.in \epsffile{Figure4_sqz.eps}
%\caption{ The maximum output squeezing for several values of $N$
%is plotted with respect to $1/N$. By extrapolation to $N=\infty$,
%we obtain the $V_{\rm min}(\omega=0)=0.05$, that is, $-13{\rm
%dB}$.} \label{Qsim3}
%\end{figure}

\clearpage

\newpage
\end{document}